\documentclass[usenatbib,iop,apj]{emulateapj}

\usepackage{natbib}
\usepackage{amsmath}
\usepackage{amssymb}
\usepackage{amsbsy}

\usepackage{ulem}

\def\eV{{\rm eV}} 
\def\MeV{{\rm M}\eV} 
\def\GeV{{\rm G}\eV} 
\def\TeV{{\rm T}\eV} 
\def\erg{{\rm erg}} 

\def\cm{{\rm cm}}
\def\s{{\rm s}}

\newcommand\bmath[1] {\mbox{\boldmath$\rm #1$}}

\def\GIC{\Gamma_{\rm IC}}

\def\Dpp{D_{\rm pp}}

\def\nb{n_{b}}
\def\nIGM{n_{\rm IGM}}

\def\bp{\bmath{p}}

\def\gph{\gamma_{\rm ph}}
\def\vph{v_{\rm ph}}

\def\cG{\mathcal{G}}

\def\gz{\gamma_{z'}}
\def\vz{v_{z'}}

\newcommand{\vel}{\ensuremath{\bmath{v}}}
\newcommand{\pmom}{\ensuremath{\bmath{p}}}
\newcommand{\Efield}{\ensuremath{\bmath{E}}}
\newcommand{\Bfield}{\ensuremath{\bmath{B}}}
\newcommand{\grad}{\ensuremath{\bmath{\nabla}}}
\newcommand{\gradp}{\ensuremath{\bmath{\nabla}_p}}
\newcommand{\ppos}{\ensuremath{\bmath{x}}}

\newcommand{\fLAB}{\ensuremath{f_{\rm L}}}
\newcommand{\fCM}{\ensuremath{f_{\rm COM}}}

\newcommand{\dVLAB}{\ensuremath{d^3p_{\rm L} d^3x_{\rm L}}}
\newcommand{\dVCM}{\ensuremath{d^3p_{\rm COM} d^3x_{\rm COM}}}

\newcommand{\betaCMpara}{\ensuremath{\beta_{\rm COM,\parallel}}}
\newcommand{\betaCMperp}{\ensuremath{\beta_{\rm COM,\perp}}}

\newcommand{\betaCM}{\ensuremath{\bmath{\beta_{\rm COM}}}}

\newcommand{\betabar}{\ensuremath{\overline{\bmath{\beta}}}}
\newcommand{\betabarLAB}{\ensuremath{\betabar_{\rm L}}}
\newcommand{\betabarCM}{\ensuremath{\betabar_{\rm COM}}}
\newcommand{\betaBOOST}{\ensuremath{\beta_b}}

\newcommand{\betabarLABpara}{\ensuremath{\overline{\beta}_{\rm L,\parallel}}}
\newcommand{\betabarLABperp}{\ensuremath{\overline{\beta}_{\rm L,\perp}}}

\newcommand{\dbetabarLABpara}{\ensuremath{\overline{\Delta\beta^2}_{\rm L, \parallel}}}
\newcommand{\dbetabarLABperp}{\ensuremath{\overline{\Delta\beta^2}_{\rm L, \perp}}}
\newcommand{\dbetabarCMi}{\ensuremath{\overline{\Delta\beta^2}_{\rm COM, i}}}
\newcommand{\dbetabarCMpara}{\ensuremath{\overline{\Delta\beta^2}_{\rm COM, \parallel}}}
\newcommand{\dbetabarCMperp}{\ensuremath{\overline{\Delta\beta^2}_{\rm COM, \perp}}}

\newcommand{\epm}{\ensuremath{e^{\pm}}}

\newcommand{\kvec}{\ensuremath{\bmath{k}}}
\newcommand{\rvec}{\ensuremath{\bmath{r}}}
\newcommand{\jvec}{\ensuremath{\bmath{j}}}

\newcommand{\xphat}{\ensuremath{\hat{\bmath{x}}'}}

\newcommand{\zphat}{\ensuremath{\hat{\bmath{z}}'}}

\def\G{{\rm G}}
\newcommand{\kI}{\ensuremath{\mathcal{I}}}

\newcommand{\gammabeam}{\ensuremath{\gamma_{b}}}
\newcommand{\vbeam}{\ensuremath{v_{b}}}

\def\Fermi{{\em Fermi\ }}

\begin{document}

\title{The Linear Instability of Dilute Ultrarelativistic \epm\ Pair Beams} 

\author{
Philip Chang\altaffilmark{1},
Avery E.~Broderick\altaffilmark{2,3},
Christoph Pfrommer\altaffilmark{4},
Ewald Puchwein\altaffilmark{5},
Astrid Lamberts\altaffilmark{6},
Mohamad Shalaby\altaffilmark{2,3,7},
Geoffrey Vasil\altaffilmark{8}
}
\altaffiltext{1}{Department of Physics, University of Wisconsin-Milwaukee, 3135 North Maryland Avenue, Milwaukee, WI 53211, USA; chang65@uwm.edu}
\altaffiltext{2}{Perimeter Institute for Theoretical Physics, 31 Caroline Street North, Waterloo, ON, N2L 2Y5, Canada}
\altaffiltext{3}{Department of Physics and Astronomy, University of Waterloo, 200 University Avenue West, Waterloo, ON, N2L 3G1, Canada}
\altaffiltext{4}{Heidelberg Institute for Theoretical Studies, Schloss-Wolfsbrunnenweg 35, D-69118 Heidelberg, Germany; christoph.pfrommer@h-its.org}
\altaffiltext{5}{Institute of Astronomy and Kavli Institute for Cosmology, University of Cambridge, Madingley Road, Cambridge, CB3 0HA, UK}
\altaffiltext{6}{TAPIR, Mailcode 350-17, California Institute of Technology, Pasadena, CA 91125, USA}
\altaffiltext{7}{Department of Physics, Faculty of Science, Cairo University, Giza 12613, Egypt}
\altaffiltext{8}{School of Mathematics \& Statistics, University of Sydney, NSW 2006, Australia}


\begin{abstract}
The annihilation of TeV photons from extragalactic TeV sources and the extragalactic background light produces ultrarelativistic $\epm$ beams, which are subject to powerful plasma instabilities that sap their kinetic energy.
Here we  study the linear phase of the plasma instabilities that these pair beams drive.  To this end, we calculate the linear growth rate of the beam plasma and oblique instability in the electrostatic approximation in both the reactive and kinetic regimes,  assuming a Maxwell-J{\"u}ttner distribution for the pair beam.  We reproduce the well-known reactive and kinetic growth rates for both the beam plasma and oblique mode. We demonstrate for the oblique instability that there is a broad spectrum of unstable modes that grow at the maximum rate for a wide range of beam temperatures and wave vector orientations relative to the beam.  We also delineate the conditions for applicability for the reactive and kinetic regimes and find that the beam plasma mode transitions to the reactive regime at a lower Lorentz factor than the oblique mode due to a combination of their different scalings and the anisotropy of the velocity dispersions.  Applying these results to the ultrarelativistic $\epm$ beams from TeV blazars, we confirm that these beams are unstable to both, the kinetic oblique mode and the reactive beam-plasma mode.  These results are important in understanding how powerful plasma instabilities may sap the energy of the ultrarelativistic $\epm$ beams as they propagate through intergalactic space.
\end{abstract}

\section{Introduction}\label{sec:introduction}

The \Fermi satellite and ground-based imaging atmospheric Cherenkov telescopes such as
H.E.S.S., MAGIC, and VERITAS\footnote{High Energy
  Stereoscopic System, Major Atmospheric Gamma Imaging Cerenkov Telescope, Very
  Energetic Radiation Imaging Telescope Array System.} have demonstrated that the high
energy Universe is teeming with energetic very high-energy gamma-ray 
(VHEGR, $E > 100\,\GeV$) sources, the extragalactic component of which mainly consists of TeV
blazars with a minority population of other sources  such as radio
and starburst galaxies. These extragalactic VHEGR emitters produce TeV photons that are greatly attenuated via annihilation upon soft photons in the extragalactic background light
(EBL) and produce pairs \citep[see, e.g.,][]{Goul-Schr:67,Sala-Stec:98,Nero-Semi:09}.

It has been assumed that these ultrarelativistic pairs produced by VHEGR annihilation 
lose energy exclusively through inverse-Compton (IC) scattering off of the cosmic microwave
background (CMB),   transferring the energy of the original VHEGR to gamma-rays with
energies $\lesssim100\,\GeV$.  The absence of observed secondary IC emission leads a number of
authors to argue that this \textit{lack} of emission places lower bounds upon the
intergalactic magnetic field
\citep[IGMF; see, e.g.,][]{Nero-Vovk:10,Tave_etal:10a,Tave_etal:10b,Derm_etal:10,Tayl-Vovk-Nero:11,Taka_etal:11,Dola_etal:11} with 
typical numbers ranging from $10^{-19}\,\G$ to $10^{-15}\,\G$.  

In addition, \Fermi has also provided the most precise estimate of the unresolved
extragalactic gamma-ray background (EGRB) for energies between
$200\,\MeV$ and $100\,\GeV$.  Since inverse-Compton cascades (ICCs) reprocess the VHEGR emission of
distant sources into this band, this has been used to constrain the
evolution of the luminosity density of VHEGR sources
\citep[see, e.g., ][]{Naru-Tota:06,Knei-Mann:08,Inou-Tota:09,Vent:10}.
These constraints preclude any dramatic rise in numbers of source by $z\approx1$--$2$ that is seen in the quasar
distribution.  That is, the comoving number of blazars must have
remained essentially fixed, at odds with both the physical picture
underlying these systems and with the observed evolution of similarly
accreting systems, i.e., quasars and radio galaxies.

These two important conclusions depend on IC cooling dominating the evolution of the ultra-relativistic pairs.  However, it was recently found that plasma instabilities driven by the
ultrarelativistic pair beams  likely are the dominant cooling mechanisms
(\citealt{paperI}, hereafter BCP12, \citealt{Schlickeiser+12,Schlickeiser+13}), depositing this energy
as heat in the intergalactic medium \citep{paperII,paperIII}.  Therefore, the lack of an observed
 IC halo emission from TeV blazars does not imply the existence of the IGMF as previous groups have argued
(BCP12;\citealt{Schlickeiser+12,Schlickeiser+13}). We note that the effectiveness of these plasma instabilities is complicated by nonlinear effects, which we briefly discuss below.

The deposition of kinetic energy into the IGM via plasma instabilities produces excess heating, which over cosmological time, may resolve a variety
of puzzles, including explaining
anomalies in the statistics of the high-redshift Ly$\alpha$ forest
\citep{paperIV,2015ApJ...811...19L} and potentially explaining a number of the X-ray
properties of groups and clusters and anomalies in galaxy formation on
the scale of dwarfs \citep{paperIII,Lu+2013}.  We
have recently shown that \textit{if} the IC halos are ignored, it is possible
to quantitatively reproduce the redshift and flux distributions
of nearby hard gamma-ray blazars and the extragalactic gamma-ray background
spectrum above 3~GeV simultaneously with a unified model of AGN evolution
\citep{Broderick+2013,Broderick+2013b}.  All of these empirical successes provide
circumstantial evidence for the presence of virulent plasma beam
instabilities. 

These potential implications of blazar heating rely on an understanding of the linear and nonlinear physics of these plasma instabilities.  Recent work in this area has been inconclusive.  For instance, \citet{Miniati+12} argued that these instabilities are physically irrelevant for the cooling of these pair beams because they would saturate at a very low level due to nonlinear Landau damping (NLD).  However, \citet{Chang+14} performed a detailed calculation of NLD to show that these plasma processes remain dominant.  In addition, \citet{Sironi+14} performed particle-in-cell simulations of these plasma processes and argued that these processes saturate at a very low level.  It is unclear, however, if the conclusions of their work is applicable to the parameter regime of blazar heating. 

Additional nonlinear effects may also be important. For instance, for sufficiently powerful blazers, the modulation instability may operate \citep{Schlickeiser+12,Chang+14,2015MNRAS.448.3405M}, allowing for a rapid transfer of electrostatic wave energy into thermal energy.  For less powerful blazars, the combination of NLD and quasilinear damping, i.e., beam plateauing, will also reduce the rate of damping compared to the linear rate, and alters the resulting IC spectra \citep{2015MNRAS.448.3405M}. Further study of these effects will help clarify these points.

While a full nonlinear study is required, we focus on the nature of the linear instability in this paper, clarifying its robustness and regimes of applicability.  We begin by studying the distribution function of the $\epm$ pairs that are produced from VHEGR-EBL photon annihilation.  We study the evolution of a distribution function under Lorentz transformations to develop an analytic understanding of how the perpendicular and parallel velocity dispersions transform under boosts.  Using this understanding, we then develop a simple description of the distribution function of the beam, which we then use to calculate the unstable modes analytically in both the reactive (hydrodynamic) and kinetic regimes.

Here the reactive instability refers to the instability where the entire beam participates in the instability.  In particular, all the beam particles are resonant with the unstable wave on a timescale longer than the growth time of the instability.  The reactive instability is also referred to as the hydrodynamic instability since the instability can be derived from the fluid equations instead of kinetic theory.  On the other hand, in the kinetic regime, only a fraction of the beam particles are resonant with the beam over a the growth time of the instability, which reduces the growth rate compared to the reactive instability for the same beam density and beam Lorentz factor.  We recover the well-known results for the reactive regime for both the beam-plasma and oblique modes.  We also derive the growth rate for these two instabilities in the kinetic regime and delineate the range of applicability for both the reactive and kinetic cases and apply these results for ultrarelativistic \epm\ pair beams. 

This paper is organized as follows.  In Section \ref{sec:setup}, we describe the transformation properties of an ultrarelativistic \epm\ beam in terms of its distribution function.  We then calculate the various linear instabilities that this beam is subject to in section \ref{sec:linear}.  In particular, we pay careful attention to both the reactive (or hydrodynamic) and kinetic regimes of the beam plasma and oblique instabilities and the transition between the two.  Applying these results to TeV $\epm$ pair beams that arise from TeV photon pair production in Section \ref{sec:application},
we demonstrate that despite the extraordinary coldness of the beam we are always in the kinetic regime for the oblique mode, but may be in the reactive regime for the beam plasma mode.  However, for the relevant parameters, the growth rates calculated in either regime are similar.  We close with a discussion of the implications of this work and application of these results for nonlinear theory in Section \ref{sec:conclusions}.

\section{Ultrarelativistic Pair Beams from VHEGRs}\label{sec:setup}

As stated in the Introduction, VHEGR photons pair produce on encountering EBL
photons as they propagate throughout the universe \citep{Gould+66}, and this
attenuation of VHEGR flux has been used as a probe of the EBL
\citep{Stec-deJa-Sala:92,deJa-Stec-Sala:94,Ahar_etal:06}.  The basic requirement
of this process is that the energies of the VHEGR ($E_{\rm ph}$) and the EBL
photon ($E_{\rm ebl}$) exceed the rest mass energy of the $\epm$ pair in the
center of momentum (COM) frame, i.e., $2 E E_{\rm ebl}(1-\cos\theta) \geq 4
m_e^2 c^4$, where $\theta$ is the relative angle of propagation in the lab
frame. As a result, an $\epm$ pair can be produced with Lorentz factor$\gamma = \left(1-v^2/c^2\right)^{-1/2} \approx E/2m_e c^2$, where $v$ is the velocity of the pairs \citep{Goul-Schr:67}.  Here, we discuss the distribution function of the
pair beam that emerges from this process.

\subsection{Distribution Function of the Pair Beam}

In the COM frame of the beam, we assume that the distribution function is isotropic, such that $f=f(E)$ is just a function of energy.   This equilibrium energy distribution of a relativistic thermal plasma gas is
\begin{equation}\label{eq:relativistic maxwellian}
f\propto \exp\left(-\frac E {k_BT}\right),
\end{equation}
 where $E$ and $T$ are the dimensionless energy and temperature in terms of a particles rest mass.  In the non-relativistic case, this reduces to the Maxwell-Boltzmann distribution, while the relativistic version is known as the Maxwell-J{\"u}ttner distribution \citep{1911AnP...340..145J}. 

The relativistic Maxwellian distribution can be extended to a drifting (or boosted) distribution via an appropriate Lorentz transformation.  The relationship between the energies of the lab (boosted) frame and the COM frame is
\begin{equation}
  E_{\rm COM} = \gammabeam\left(E_{\rm L} - \betaBOOST p_{\rm L,\parallel}\right),
\end{equation}
where $\gammabeam = \gamma(v_b) = \left(1 - v_b^2/c^2\right)$ is the Lorentz factor of the beam and $v_b$ is the bulk velocity of the pair beam.
Inserting this into (\ref{eq:relativistic maxwellian}), we find the Maxwell-J{\"u}ttner distribution \citep{1911AnP...340..145J,1975PhRvA..12..686W}
\begin{eqnarray} 
f &=& \frac{n_b m_e c^2}{4\pi\gammabeam k_B T_b K_2(m_ec^2/k_B T_b )m_e^3c^3} \nonumber\\
&&\times \exp\left(-\frac {\gammabeam(E - \vel_b\cdot \bmath{p})} {k_B T_b}\right),\label{eq:distribution function beam}
\end{eqnarray}
where $K_2$ is the 2nd order modified Bessel function and $T_b$ is the comoving temperature of the beam.

The Maxwell-J{\"u}ttner distribution leads to an anisotropic velocity spread parallel and perpendicular to the beam's direction.  In Appendix \ref{sec:lorentz}, we estimate how the parallel and perpendicular velocity spreads scale.  The relevant results are: 
\begin{equation}\label{eq:perp dispersion}
\frac{\Delta v_{\perp}^2}{c^2} \approx \frac {2 k_B T_b}{\gammabeam^2 m_e c^2 }
\quad\text{and}\quad
\frac{\Delta v_{\parallel}^2}{c^2} \approx \frac {k_B T_b}{\gammabeam^4 m_e c^2 },
\end{equation}
where $T$ is measured in the COM frame of the beam.  
These simple scalings of the perpendicular velocity dispersion and parallel velocity dispersions can be understood as a results of time dilation between two frames that are boosted relative to each other, giving one factor of $\gamma^{-1}$. The coordinates perpendicular to the boost axis remain invariant while the axis along the boost suffers from length contraction and gives an extra scaling of $\gamma^{-1}$ for the parallel case.  In any case, an ultrarelativistic beam has a small velocity spread in both the parallel and perpendicular directions by factors of $\gamma^{-2}$ and $\gamma^{-1}$, respectively.  These velocity dispersions will be important in delineating the regime of instability in \S~\ref{sec:transition}. 

While we have modeled the pair distribution function as a Maxwell-J{\"u}ttner distribution, the physical distribution function that is produced by VHEGR photo annihilation is somewhat more complicated \citep[see for instance][]{2012ApJ...758..101S}.  In particular, the parallel and perpendicular momentum spread will be influenced by the distribution of VHEGR photons and their respective mean free paths.  However, a Maxwell-J{\"u}ttner distribution is still useful. First, it also is sufficiently simple to allows us to calculate the kinetic instability exactly in the electrostatic approximation.  Second, its instability growth rates has been calculated without approximation previously by \citet{Bret-Grem-Beni:10}, allowing a point of comparison for our calculation using the electrostatic approximation (as mentioned below).  Third, it possess a continuous (small) distribution of parallel and perpendicular momenta that allow us to elucidate the physics.   Finally, the analytic methodology used to calculate the Maxwell-J{\"u}ttner distribution may be useful for the full calculation using the physical distribution function.

\section{Linear Theory}\label{sec:linear}

The Vlasov equation for each species is
\begin{equation}\label{eq:linearize vlasov}
  \frac {\partial f_s}{\partial t} + \vel_s \cdot \grad f_s + 
  q_s \left({\Efield} + \frac {\vel_s} {c} \times {\Bfield} \right)
 \cdot\grad_p f_s = 0,
\end{equation}
where $\vel_s=\pmom_s/\gamma_s m_e$ and  $\gamma_s = 1/\sqrt{1-v_s^2/c^2} $
Here, $s$ is the species label, $+$ for positrons and $-$ for electrons,
with $q_{\pm} = \pm e$. 
Upon linearizing this in small perturbations about a background distribution, i.e., setting 
$f_s\rightarrow f_{0s} + \delta f_s$, $\Bfield \rightarrow \delta \Bfield$ and $\Efield \rightarrow \delta \Efield$,  we obtain, 
\begin{equation}\label{eq:vlasov}
  \frac {\partial\delta f_s}{\partial t} + {\vel_s} \cdot \grad\delta f_s  
  +  {q_s} \left(\delta \Efield + \frac {\vel_s} {c} \times \delta \Bfield \right)\cdot\grad_p f_{0s} = 0.
 \end{equation}
The plasma couples to the field through the Maxwell equations
\begin{eqnarray}\label{eq:maxwell}
\grad \times \delta \Efield &=& -\frac 1 c \frac {\partial \delta \Bfield} {\partial t},\\
\grad \times \delta \Bfield &=& \frac {4\pi} {c} \delta\jvec + \frac{1}{ c} \frac {\partial \delta \Efield} {\partial t},\label{eq:maxwell2}
\end{eqnarray}
where $\delta\jvec = \sum_s q_s\int \vel \delta f_s d^3 p$ is the linear current density perturbation.

Here it is useful to work within the electrostatic approximation ($\kvec \times \delta \Efield = 0$), where we only need to include Coulomb's law for the electric field rather than the full Maxwell equations:
\begin{equation}
 i\kvec\cdot\delta\Efield = 4\pi\delta\rho,
\end{equation}
where $\delta\rho = \sum_s q_s\int \delta f_s \,d^3\!p$ is the perturbed charge density. 
By adopting the electrostatic approximation, we have explicitly ignored electromagnetic modes.  This would preclude, for example, the Weibel instability.
In addition, the electromagnetic terms would introduce corrections to the physics that are not necessarily small in the limit of relativistic particles, i.e., $v/c \rightarrow 1$.  However, we make this approximation for two reasons.  First, a complete calculation of the unstable modes has already been carried out by \citet{Bret-Grem-Beni:10}, who showed that the oblique mode is mainly electrostatic (modulo the Weibel instability). Hence   a electrostatic approximation to the full dispersion relation should recover the essential physics.  Second, the electrostatic approximation is much simpler than a full calculation and allows us to analytically calculate the unstable growth rates, while permitting a clear exposition of the relevant physics. 

We now adopt perturbations of the form  $\delta \propto \exp\left(i \kvec\cdot\rvec - i \omega t\right)$ and without loss of generality assume that $\kvec = (k_x, 0, k_z)$, where $k_z$ is along the beam direction.  Linearizing the Vlasov-Maxwell equations then leads to the dispersion relation:
\begin{equation}\label{eq:dispersion relation}
 \epsilon = 1 + \sum_s \frac{m_e \omega_{p,s}^2}{k^2}\int \frac{\kvec \cdot \gradp F_s}{\omega - \kvec\cdot\vel} d^3\!p = 0,
\end{equation}
where $\epsilon$ is the simplified dielectric function, and for each species $\omega_{p,s}^2\equiv 4\pi e^2 n_s/m_e$ is the plasma frequency, $n_s\equiv\int f_{0s} d^3\!p$ is the number density, and $F_s\equiv f_{0s} / n_s$ is the normalized background distribution function.
Upon integrating by parts, Equation (\ref{eq:dispersion relation}) becomes
\begin{multline}
  \epsilon = 
  1 - \sum_s \frac{m_e\omega_{p,s}^2}{k^2} \int F_s\kvec \cdot \gradp\frac{1}{\omega - \kvec\cdot\vel} d^3p\\ 
  =
  1 - \sum_s \frac{\omega_{p,s}^2}{k^2c^2}\int F_s \frac{k^2c^2 - (\kvec\cdot\vel)^2}{\gamma(\omega - \kvec\cdot\vel)^2} d^3p 
  =
  0\,.
  \label{eq:dispersion relation by parts}
\end{multline}

There are two distinct, often qualitatively different, regimes in which we may
consider the implications of this dispersion relation.  The first is the cold
plasma limit or the hydrodynamic or reactive limit.  The hydrodynamic limit is
aptly named because the resulting dispersion relation that is found could have
also been calculated directly from the continuity equation and the momentum
equation.  In this limit, the internal distribution of the particles of the
background or beam are irrelevant to the physics of the instability and it is
only the bulk response that is important.  In particular, this means that the
beam particles are resonant with the unstable wave over a timescale much longer
than the growth time, i.e., the beam particles
do not drift a distance larger than the wavelength of the unstable mode over the
growth time of the instability.  The second is the kinetic regime, where the
internal distribution of  beam particles is
important to the physics of the instability. Here, only a fraction of
 beam particles stay within one wavelength of the unstable mode
over the growth time of the instability.  Moreover, the bulk of
the plasma (background or beam) does not respond to the disturbance; instead,
only a fraction of  particles is relevant for
driving (instability) or damping (Landau damping).  We discuss below the
evaluation of the dispersion relation in these two regimes, which gives two
regimes of instability, and the delineation between them. 

\subsection{Hydrodynamic (Reactive) Instability}

Starting with the dispersion relation (\ref{eq:dispersion relation by parts}), we first consider the instability of a cold plasma beam. 
Taking the limit of Equation (\ref{eq:distribution function beam}) as $k_BT_t  \rightarrow 0$, for a target plasma $v_0=0$ and a beam plasma $v_0=\vbeam$\footnote{That is, we set $F_s(\bmath{p})=\delta^3(\bmath{p}-\bmath{p}_{0s})$ where $\bmath{p}_{0s}\equiv \gamma_0 m_e v_0 \hat{\bmath{z}}$ is the momentum associated with $v_0$.}, we find
\begin{equation}\label{eq:dispersion relation reactive}
 1 - \frac{\omega_{p,t}^2}{\omega^2} - \frac{\omega_{p,b}^2}{\gamma^3(\omega-k_z \vbeam)^2}\frac{\gamma^2 k_x^2 + k_z^2}{k_x^2 + k_z^2} = 0.
\end{equation}
For $k_x = 0$, we recover the same beam-plasma instability which was described in the Appendix of BCP12.  

The solution to Equation (\ref{eq:dispersion relation reactive}) is given in Appendix \ref{sec:solution reactive} where we show that the associated growth rate (Equation \ref{eq:growth rate reactive appendix}) is 
\begin{equation}\label{eq:growth rate reactive}
 \Gamma = \frac{\sqrt{3}}{2^{4/3}}\left(\frac{n_b}{n_t}\right)^{1/3}\left(\frac{\gamma^2 Z_x^2 + 1}{Z_x^2 + 1}\right)^{1/3}\frac{\omega_{p,t}}{\gamma},
\end{equation}
where $Z_x = k_x\vbeam/\omega_{p,t}$ is the dimensionless wavevector perpendicular to the beam direction.

For $k_x = 0 \rightarrow Z_x = 0$, this reduces to the beam-plasma growth rate, which is
\begin{equation}\label{eq:growth rate bp}
 \Gamma = \Gamma_{\rm TS} \equiv \frac{\sqrt{3}}{2^{4/3}}\left(\frac{n_b}{n_t}\right)^{1/3}\frac{\omega_{p,t}}{\gamma},
\end{equation}
which we denote the beam-plasma or ``two-stream'' growth rate.
For the more general case where $Z_x\ne 0$, this becomes the oblique instability studied by \citet{Bret-Grem-Beni:10}  Indeed for $\gamma \gg 1$ and $Z_x\gg1$, the growth rate approaches the oblique growth rate:
\begin{equation}
 \Gamma = \Gamma_{\rm ob} \equiv \frac{\sqrt{3}}{2^{4/3}}\left(\frac{n_b}{n_t}\right)^{1/3}\frac{\omega_{p,t}}{\gamma^{1/3}},
\end{equation}
which is much faster than the beam-plasma growth rate, $\Gamma_{\rm TS}$

We should caution in the derivation above that the resonance condition, which is $\omega_{p,t} - k_z \vbeam$, implies that $k_z \neq 0$.  For the case where $k_z \rightarrow 0$, the electrostatic approximation no longer holds and the full dispersion relation must be solved.\footnote{We thank Antoine Bret for helping to clarify this point.}  A solution to the full dispersion relation reveals additional modes, including the zero frequency ($k_z = 0$) filamentation or Weibel mode. 

\begin{figure}
\includegraphics[width=0.5\textwidth]{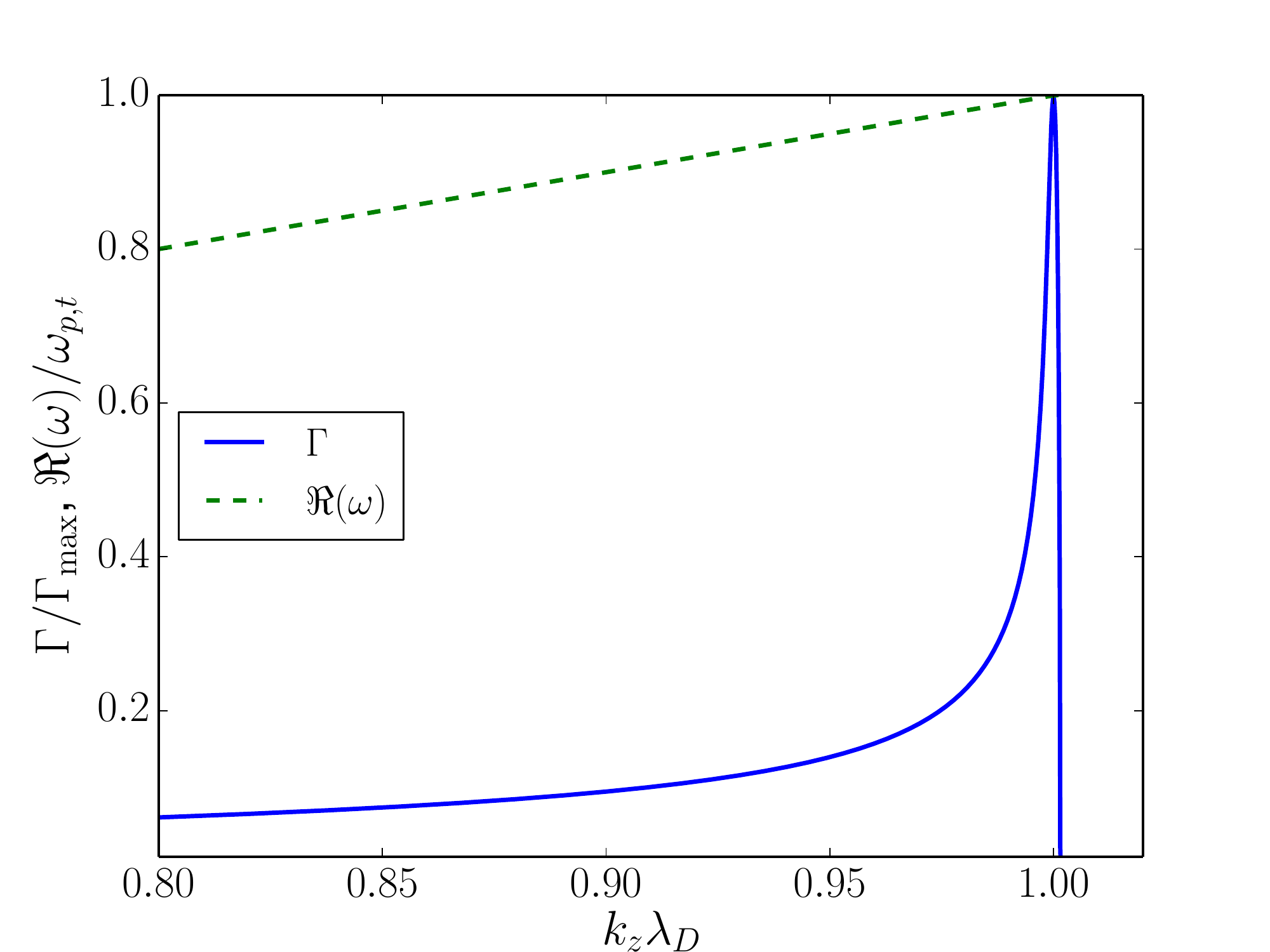}
\caption{Beam-plasma growth rate (solid line) and unstable wave frequency (dashed-line) as a function of
  $k_z$ for $\gammabeam = 100$ and $n_b/n_t=10^{-3}$. \label{fig:OffResonance}}
\end{figure}

Equation (\ref{eq:dispersion relation reactive}) can also be solved numerically in terms of $k_x$ and $k_z$.  Here let us specialize to the case of $k_x = 0$, i.e., the beam-plasma case.  In this case, we have 
\begin{equation}\label{eq:dispersion relation beam-plasma}
 1 - \frac{\omega_{p,t}^2}{\omega^2} - \frac{\omega_{p,b}^2}{\gamma^3(\omega-k_z \vbeam)^2} = 0,
\end{equation}
which we can numerically solve in terms of $\omega/\omega_{p,t}$, $k_z\lambda_D$, $\vbeam/c$, and $n_b/n_t$, where $\lambda_D = c/\omega_{p,t}$ is the skin depth.  In Figure \ref{fig:OffResonance} we show the real and imaginary parts for $\omega/\omega_{p,t}$ as a function of $k_z\lambda_D$ for the representative case of $\vbeam/c \approx 1$ and $n_b/n_t = 10^{-3}$ and $\gamma=100$.    For $k_z\lambda_D = 1$, the growth rate reaches it maximum of $\Gamma_{\rm max}$ and the real part of the frequency is $\Re(\omega) = \omega_{p,t}$, which is the plasma oscillation frequency.  This wave would exist in the absence of a tenuous beam.  However, as we move away from this frequency toward lower $k_z$, we still find substantial growth, with $\Gamma \approx 0.1 \Gamma_{\rm max}$ as $k_z \approx 0.9\lambda_D^{-1}$.  Interestingly, the real part of the unstable wave has a phase velocity, $v_{\rm ph} = \Re(\omega)/k = c$, which is still in resonance with the beam.  

In a continuous system, these waves do not matter in comparison to the unstable mode at $k\lambda_D = 1$.   However, for discrete numerical systems, which do not sufficiently resolve the most unstable modes, these sub-maximal modes drive the growth of the instability of numerically calculated beam-plasma systems, which may lead to an incorrect nonlinear state in comparison to the physical system.

\subsection{Kinetic Instability}

The growth rate expressed in Equation (\ref{eq:growth rate reactive}) is in the reactive (or hydrodynamic) regime as the dispersion relation (Equation \ref{eq:dispersion relation reactive}) could have been derived from the fluid equations.  Here all the particles participate in the instability.  However, kinetic theory marks another regime of the instability, where only a fraction of the particle participate in the instability, i.e., the kinetic regime.  We now derive the growth rate of the instability in the kinetic regime.

We begin first with the distribution function for the target plasma:
\begin{equation}
 F_t = \left(\frac{1}{2\pi m_e k_B T_t}\right)^{3/2}\exp\left(-\frac{p^2}{2m_e k_B T_t}\right),
\end{equation}
where the target plasma is assumed to be nonrelativistic, $\bmath{p}=m_e\vel$ is the nonrelativistic momentum,  and $T_t$ is the temperature of the target background plasma.
For the beam plasma, we again adopt the Maxwell-J{\"u}ttner distribution (Equation~\ref{eq:distribution function beam}).
Inserting these into the dispersion relation (Equation \ref{eq:dispersion relation}), we find
\begin{eqnarray}
 1 &-& \frac{\omega_{p,t}^2}{k^2c^2}\int F_t\frac{k^2c^2 - (\kvec\cdot\vel)^2}{\gamma(\omega - \kvec\cdot\vel)^2} d^3p \nonumber \\
 &+& \frac{m_e\omega_{p,b}^2}{k^2}\int \frac{\kvec\cdot\gradp F_b}{\omega - \kvec\cdot\vel}d^3p  = 0,\label{eq:dispersion kinetic}
\end{eqnarray}
where we have integrated by parts only the second term, associated with the target plasma. 

We discuss the solution to Equation (\ref{eq:dispersion kinetic}) in Appendix \ref{sec:solution kinetic}.  The associate growth rate for the kinetic oblique instability is  
\begin{equation}
\begin{aligned}
\Gamma &\approx 
- \Gamma_0
\frac{\pi \gamma_w^2 \gph^3 (\vph-v_{b,z'})}{4 \gammabeam \mu^2 K_2(\mu) \cG'^3 c}\\
&\times\left[
\left( \cG'^2\mu^2 + 2\cG'\mu + 2 \right) 
+
\frac{\gammabeam^2 v_{b,x'}^2}{2\cG'^2c^2} \left(2 \cG'\mu+ 2\right)
\right]
\exp(-\cG'\mu),\label{eq:kinetic growth rate}
\end{aligned}
\end{equation}
where $\mu = m_e c^2/k_B T_b$, $\gph = \left(1 - \vph^2/c^2\right)^{-1/2}$, $\vph = \omega/k$ is the phase velocity of the wave, $v_{b,z'}$ is the velocity of the beam oriented along the wavevector, $v_{b,x'}$ is the velocity of the beam perpendicular to the wavevector, $\gamma_w = \left(1 - w^2/c^2\right)^{-1/2}$, $w = \gamma_{\rm ph}^{-1}v_{b,x'}/(1 - v_{b,z'}v_{\rm ph}/c^2)$ is the beam velocity transverse to the wavevector in a frame that is comoving with the wave at its phase velocity, and $\cG'\equiv\gammabeam\gph(1-v_{b,z'}\vph/c^2)/\gamma_w$ is the Lorentz factor of a beam particle in a frame that is comoving with the wave at the phase velocity and the transverse (to the wavevector) beam bulk velocity 
Finally, we define the typical maximum growth rate, $\Gamma_0$, as 
\begin{equation}
\Gamma_0 \equiv \omega_p \gammabeam\frac{n_b}{n_t} \frac{m_e v_b^2}{k_B T_b}. 
\end{equation}
Equation (\ref{eq:kinetic growth rate}) specializes to the beam-plasma growth rate if we take $v_{b,x'} = 0$, which gives:
\begin{equation}
\Gamma_{\rm bp} \approx 
- \Gamma_0
\frac{\pi \gph^3 (\vph-v_{b,z'})}{4 \gammabeam \mu^2 K_2(\mu) \cG^3 c}
\left( \cG^2\mu^2 + 2\cG\mu + 2 \right) 
\exp(-\cG\mu),\label{eq:bp kinetic growth rate}
\end{equation}
where we have used the fact that $\gamma_w = 1$ for $v_{b,x'} = 0$

\begin{figure}
\includegraphics[width=0.5\textwidth]{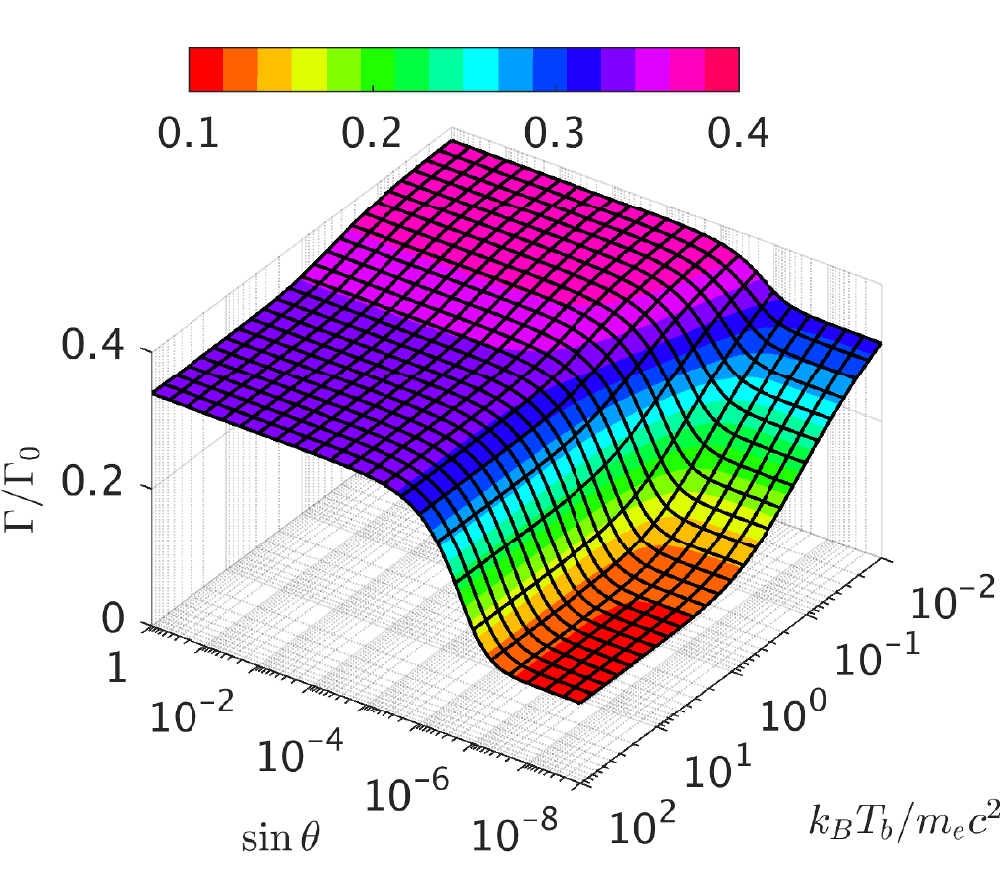}
\caption{Oblique growth rate maximized over $\vph$ as a function of
 $\sin\theta$ and $k_BT_b/m_e c^2$, where $\theta$ is the angle between the beam direction and wavevector for $\gammabeam = 10^6$. The maximum growth rate occurs when
  $\sin\theta \gg1/\gammabeam$. Note that as $\sin\theta\rightarrow 0$, we asymptote to the beam-plasma growth rate.}\label{fig:ObliqueGMZ}
\end{figure}\

In Figure \ref{fig:ObliqueGMZ}, we plot the growth rate for the oblique instability (Equation (\ref{eq:kinetic growth rate})) as a function of
$\sin\theta$, where $\cos\theta = \hat{\kvec}\cdot\hat{\vel}$, i.e., the angle between the
beam and the wavevector, and $k_BT_b/m_e c^2$.  Here, it is clear that the growth
rate reaches its maximum value $\sin\theta \gtrsim1/\gammabeam$, i.e., at an oblique angle.  Note that as $\sin\theta \rightarrow 0$, we recover the beam-plasma instability.  Moreover, the maximal growth rates, normalized to $\Gamma_0$, vary little and are
robust for a broad range of angles between the wavevector and the beam direction.  It is clear from this plot that for nearly any combination of wavevector orientation and beam temperature that there exists a broad spectrum of modes that are unstable and grow at nearly the maximum growth rate, $\Gamma_0$, for the parameters of the system, $T_b$, $\gamma_b$, and $n_b/n_t$, modulo a factor of order unity.

This does not imply that any individual mode, i.e., a mode with a fixed wavevector, will grow robustly.  Any individual mode only grows when the phase velcocity of the mode in the direction of the beam are in resonance and this resonant width is narrow.  However, the growth is robust as for any combination of wavevector orientation and beam temperature, there exists some mode that will grow at the maximum rate. 

\begin{figure*}
\includegraphics[width=0.5\textwidth]{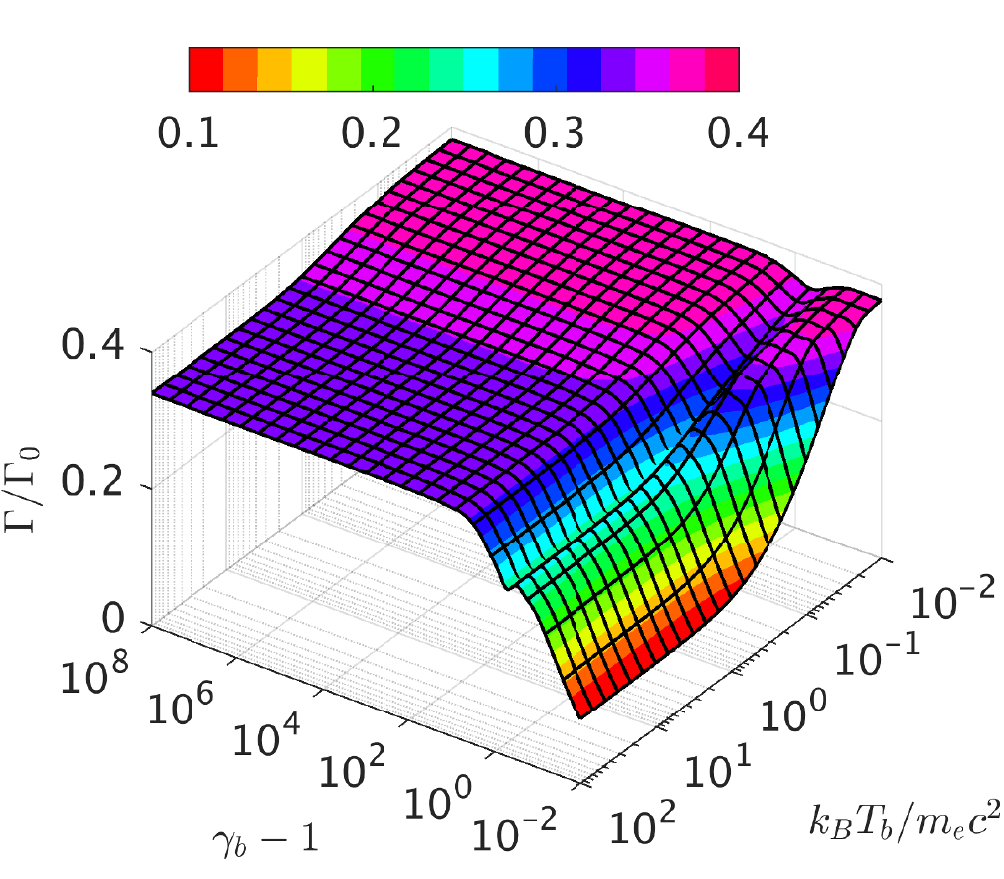}
\includegraphics[width=0.5\textwidth]{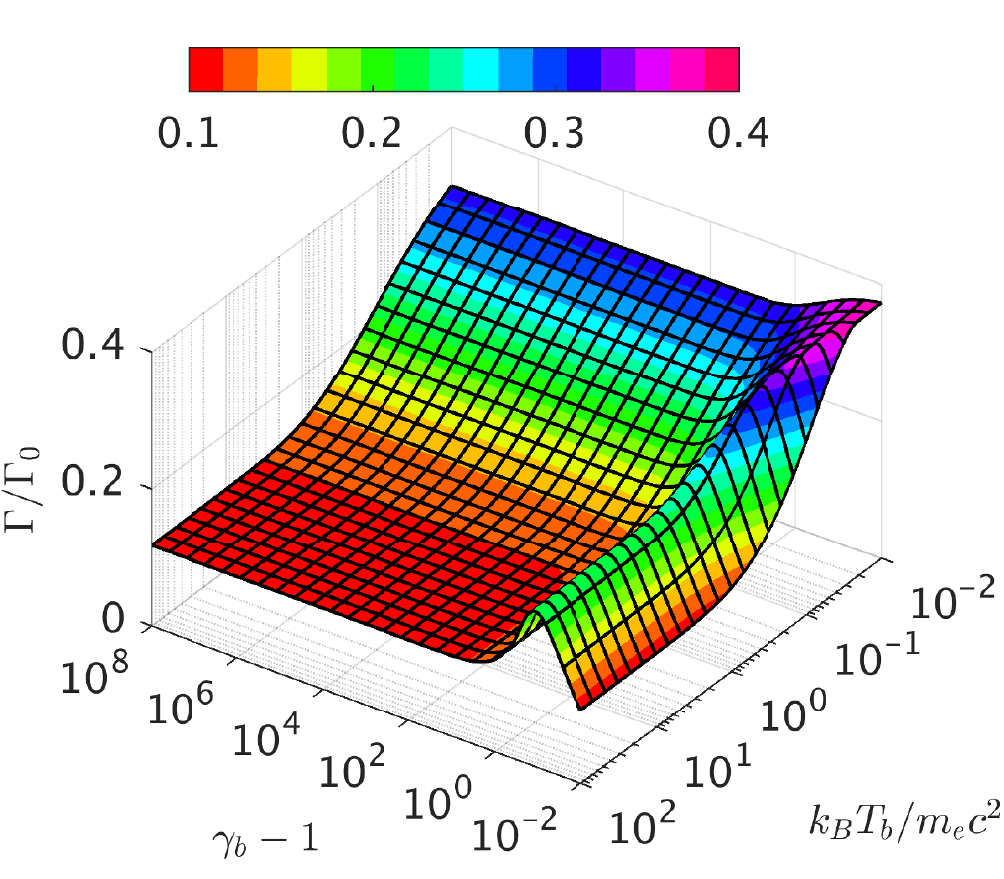}
\caption{Oblique kinetic growth rate (left) and beam-plasma growth rate (right),
  and normalized by
  $\Gamma_0\equiv\omega_{Pt}(n_b/n_t)\gammabeam m \vbeam^2/k_BT_b$ as a function of
  $k_BT_b/m_ec^2$ and $\gammabeam-1$.  The oblique kinetic growth rate has been maximized over $\theta$. Unlike the beam-plasma case on the right, in the case of the
    oblique kinetic growth rate at high $\gammabeam$ the transition between high
    and low temperature is only marginal, constituting a roughly $10\%$
    reduction. } \label{fig:OGgen}
\end{figure*}

Because there is little variation in the maximal oblique growth rate as a wavevector orientation, we plot the maximum growth rate as a function of $T_b$
and $\gammabeam - 1$ in Figure \ref{fig:OGgen}.  Here for relativistic beams, the maximum growth rate varies
little with $T_b$, varying by less than $10\%$ between hot and cold beams and 
we find:
\begin{equation}
\Gamma_M
\approx
\left\{
\begin{aligned}
& 0.38 \Gamma_0 & k_BT_b/m_ec^2\ll 1\\
& 0.34 \Gamma_0 & k_BT_b/m_ec^2\gg 1
\end{aligned}
\right.\,,\label{eq:max kinetic growth rate}
\end{equation}
for $\gammabeam\gtrsim10$, as seen in Figure \ref{fig:OGgen}.  Hence, unstable modes exists and robustly grow at roughly $\Gamma \approx 0.4\Gamma_0$ for nearly any value of $k_BT_b/m_ec^2$, wavevector orientation, and $\gammabeam \gg 1$. 

This can be contrasted with the right panel of
  Figure~\ref{fig:OGgen} where we plot the beam-plasma growth rate
(\ref{eq:bp kinetic growth rate}) as a function of $\gammabeam - 1$ and $k_B
T_b/m_ec^2$.  Here, we see that the maximum growth rate is somewhat more
sensitive to temperature, varying from $\Gamma/\Gamma_0 \approx 0.4$ for
$k_BT_b/m_ec^2 \ll 1$ to $\Gamma/\Gamma_0 \approx 0.1$ for $k_BT_b/m_ec^2 \gtrsim
1$.  Note, however, that that beam-plasma growth rate remains competitive with
the oblique growth rate, i.e., it is not orders of magnitude lower.

Finally, the maximum growth rate that we derived here (Equation (\ref{eq:max kinetic growth rate})) and that of BCP12 (their equation (16)) which is originally derived from the numerical fit of \citet{Bret-Grem-Beni:10} are exactly the same.  We must note, however, that the our definition of $T_b$ is in the COM frame whereas BCP12 defines $T_b$ in the ``lab'' frame.  As a result, there is a factor of $\gamma_b$ that is explicit in Equation (\ref{eq:max kinetic growth rate}) that is implicit in Equation (16) of BCP12.   

\subsection{The Transition between the Kinetic and Hydrodynamic Instability}\label{sec:transition}

The oblique instability exists in two different regimes, raising the important question: how are the two regimes related to each other. While this question has been studied by many authors in the context of the beam-plasma or two-stream instability \citep[see for instance][]{Melrose86,Boyd}, a clear exposition of how these two regimes are related to each other for the oblique instability is lacking.  

To begin let us return to the reactive instability.  For the growth rate of the reactive instability in Equation (\ref{eq:growth rate reactive}) to be valid, the velocity dispersion must be vanishingly small. In particular, over the growth time of the unstable wave, the beam particles may not be spread significantly, i.e., their spread is much smaller than one wavelength.  Quantitatively, this demands
\begin{equation}
 \left|\frac{\kvec\cdot\Delta \vel}{\Gamma}\right| \ll 1.
\end{equation}
For $k_\perp \approx \omega_p/v_b$, this gives
\begin{equation}\label{eq:oblique reactive regime}
 \frac{\Delta v_{\perp}}{\vbeam} \ll \left(\frac{n_b}{\gammabeam n_t}\right)^{1/3}
\end{equation}
where we have dropped constant factors of order unity and assumed that $Z_x \propto O(1)$ and that the velocity dispersion is dominated by perpendicular (to the beam) component.  Hence, this defines the upper limit on the velocity dispersion of the plasma for the cold-plasma approximation to hold and, hence, the range of validity for the reactive oblique growth rate Equation (\ref{eq:growth rate reactive}).  For $Z_x \ll \gamma^{-2}$, we recover the condition for the relativistic, reactive beam-plasma instability:
\begin{equation}\label{eq:bp reactive regime}
\frac{\Delta v_{\parallel}}{\vbeam} \ll \gammabeam^{-1}\left(\frac{n_b}{n_t}\right)^{1/3}.
\end{equation}

Applying the scaling of the perpendicular and parallel velocity dispersions (Equations \ref{eq:perp dispersion}) to these results and assuming $v_b \approx c$, we find

\begin{equation}
  \frac{\Delta v_{\perp}}{\vbeam} \approx  \gammabeam^{-1}\sqrt{\frac{k_BT_b}{m_ec^2}}
  \quad\text{and}\quad
\frac{\Delta v_{\parallel}}{\vbeam} \approx \gammabeam^{-2}\sqrt{\frac{k_BT_b}{m_ec^2}}.  
\end{equation}
Hence, the conditions for the reactive regime for the oblique mode (Equation~\ref{eq:oblique reactive regime}) and beam-plasma mode (Equation~\ref{eq:bp reactive regime}) can be reduced to 
\begin{equation}\label{eq:generic reactive regime}
 \displaystyle
1 \ll \left\{
 \begin{array}{cl}
 ({k_BT_b}/{m_ec^2})^{-1/2}\gammabeam^{2/3}\left({n_b}/{n_t}\right)^{1/3} & \textrm{oblique} \\
({k_BT_b}/{m_ec^2})^{-1/2}\gammabeam\left({n_b}/{n_t}\right)^{1/3} & \textrm{beam plasma} 
\end{array}
\right..
\end{equation}

We now proceed to study the range of validity for the kinetic growth rate for the beam plasma mode (Equation~\ref{eq:bp kinetic growth rate}) and oblique mode (Equation~\ref{eq:kinetic growth rate}).  Following the argument of \cite{Boyd}, the growth occurs over a range where the distribution function is positive or $\vbeam - \Delta v < \omega/k < \vbeam$.  Hence the bandwidth over which the distribution powers grows is  $\Delta \omega \approx k \Delta v$.  For the kinetic growth rate to be valid, the bandwidth, $\Delta\omega$, must be large compared to the growth rate; otherwise, the entire beam contributes to the growth and, hence, the reactive regime applies. For the beam plasma case, the growth rate is roughly
\begin{equation}
\Gamma \approx
\frac{n_b}{\gamma^3n_t} \left(\frac{c}{\Delta v_{\parallel}}\right)^{2}.
\end{equation}
The bandwidth, $\Delta \omega$, is then greater than the growth rate if 
\begin{equation}\label{eq:bp kinetic regime}
 \frac{\Delta v_{\parallel}}{\vbeam} \gtrsim \gamma^{-1}\left(\frac{n_b}{n_t}\right)^{1/3},
\end{equation}
which connects with the condition on the reactive beam plasma instability from Equation (\ref{eq:bp reactive regime}).
Similarly for the oblique mode, the bandwidth, $\Delta \omega$, is then greater than the growth rate if 
\begin{equation}\label{eq:oblique kinetic regime}
\frac{\Delta v_{\perp}}{\vbeam} \gtrsim \left(\frac{n_b}{\gamma n_t}\right)^{1/3},
\end{equation}
which can similarly compared to the condition on the reactive oblique mode from Equation (\ref{eq:oblique reactive regime}).

Combining these two kinetic condition and our result again from Section \ref{sec:setup}, we find  
\begin{equation}\label{eq:generic kinetic regime}
1 \ge \left\{
\begin{array}{cl}
({k_BT_b}/{m_ec^2})^{-1/2}\gammabeam^{2/3}\left({n_b}/{n_t}\right)^{1/3} & \textrm{oblique} \\
({k_BT_b}/{m_ec^2})^{-1/2}\gammabeam\left({n_b}/{n_t}\right)^{1/3} & \textrm{beam plasma} 
\end{array}
\right.,
\end{equation}
which in combination with Equation (\ref{eq:generic reactive regime}) denotes the transition between the reactive and kinetic regimes.

\section{Application to Ultrarelativistic $\epm$ Beams}\label{sec:application}

As discussed in the Introduction, the annihilation of VHEGRs and EBL photons produce ultrarelativistic \epm\ beams that are unstable to the beam plasma and oblique modes discussed above.  To apply the above results to the ultrarelativistic \epm\ beams, we now calculate their initial conditions.

\subsection{Average COM Energy of the \epm\ Beam}\label{sec:temperature}

To find the effective velocity dispersion of the ultrarelativistic \epm\ beam, we must first estimate the average COM energy of the beam.  To do so, we consider the process of photon-photon annihilation.  For a monoenergetic population of VHEGR photons with energy $E_{\rm ph}$, the angle-averaged production rate of \epm\ on EBL photons is 
\begin{equation}
\begin{split}
\Gamma_{\pm}(E_{\rm ph}) &= \frac{1}{4\pi}\int \sigma c dn_{\rm EBL}d\Omega  \\
&= \frac{1}{2}\int \sigma\left(E_{\rm ph}, E_{\rm EBL}, \theta\right) c \frac{dn_{\rm EBL}}{dE_{\rm EBL}} \\
&dE_{\rm EBL} d\cos\theta,
\end{split}
\end{equation}
where $\Gamma_{\pm}$ is the rate of pair production, $\sigma$ is the pair-production cross section, $n_{\rm EBL}$ is the number density of EBL photons, $E_{\rm EBL}$ is the energy of the EBL photons, and $\theta$ is the angle between the momentum of the VHEGR photon and the EBL photon.  There are two important components to this calculation -- the cross section, $\sigma$, and the spectrum of the EBL.  

For $\sigma$, we use the results from \citet{Nikishov62} and \citet{Gould+67}, who considered a high energy photon with energy $E_{\rm ph}$ moving along the x-axis annihilating on an EBL photon with energy $E_{\rm EBL}$ moving at an angle, $\theta$, with respect to the x-axis.  The total cross section for this process is \citep{Nikishov62,Gould+67}
\begin{equation}
\begin{split}
 \sigma = \frac {1}{2} \pi r_e^2\left(1-\frac{v_e^2}{c^2}\right)\left[\left(3-(v_e/c)^4\right)\ln\frac{1+v_e/c}{1-v_e/c} \right.\\
  \left.- 2\frac{v_e}{c}\left(2-\frac{v_e^2}{c^2}\right)\right],\label{eq:cross section}
\end{split}
 \end{equation}
where $r_e = e^2/m_e c^2$ is the classical electron radius and $v_e$ is the electron velocity in the COM frame of the generated pair.\footnote{In this section, the COM frame and subscript ``COM'' refer to the center of momentum frame of the pair that is produced by a single $\gamma-\gamma$ annihilation.}  To find $v_e$, we use the energy of the electron in the COM frame, $E_{\rm e, COM}$, which is 
\begin{equation}
\label{eq:threshold}
 E_{\rm e, COM} =  \frac{m_e c^2}{\sqrt{1-v_e^2/c^2}} = \sqrt{\frac{1}{2}{E_{\rm EBL} E_{\rm ph}\left(1-\cos\theta\right)}}.
\end{equation}
Pair production occurs when $E_{\rm e, COM}/m_e c^2 \geq 1$.

The second ingredient is the spectrum of the EBL, which is not well constrained.  Here we use the constraints from \citet{Ahar_etal:06}, who demonstrated that VHEGR emission from H 2356-309 and 1ES 1101-232 places an upper limit on the EBL that is close to the lower limit of the integrated light from galaxies \citet{Madau+00}. Looking at Figure 1 of \citet{Ahar_etal:06}, we note that the EBL has a flat spectrum, i.e., constant ${dn_{\rm EBL}}/{dE_{\rm EBL}}$ below $1\,\eV$ and a falling spectrum  ${dn_{\rm EBL}}/{dE_{\rm EBL}}\propto E_{\rm EBL}^{-1.5}$ with a spectral index of $\approx 1.5$ above $1\,\eV$ with a rapid cutoff above $10\,\eV$. Thus, we adopt a simplified model:
\begin{equation}
  \frac{dn_{\rm EBL}}{dE_{\rm EBL}} \propto 
  \begin{cases}
    E_{\rm EBL}^0 & E_{\rm EBL} \leq 1\,\eV \\
    E_{\rm EBL}^{-1.5} & 1\,\eV < E_{\rm EBL} \leq 10\,\eV \\
    0 & E_{\rm EBL} > 10\,\eV. 
  \end{cases}
\end{equation}

\begin{figure}
 \includegraphics[width=0.5\textwidth]{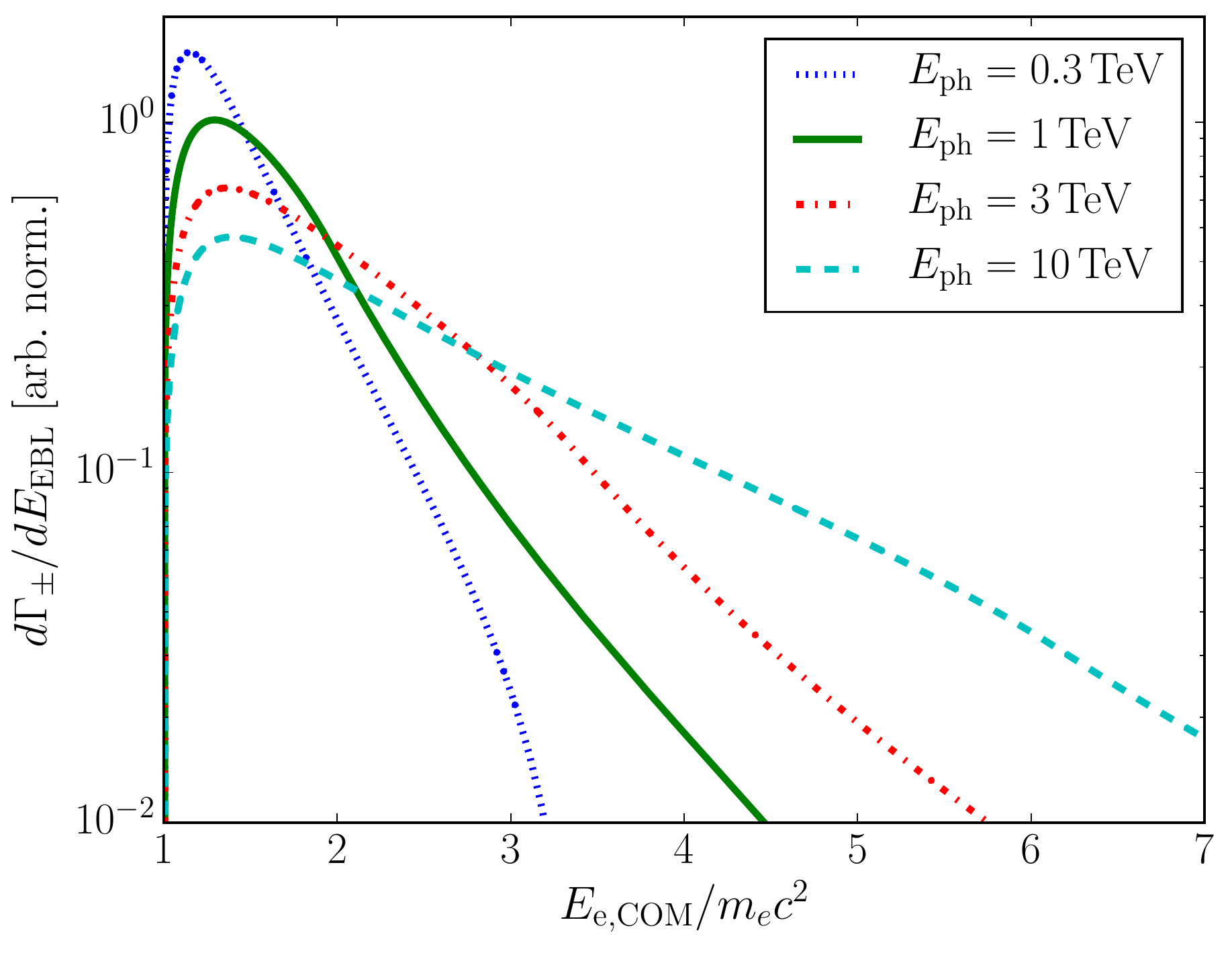}
 \caption{Differential rate of pair production as a function of the COM energy of the electron (and positron) for a photon energy of $E_{\rm ph} = 0.3$ (dotted line), $1$ (solid line), $3$ (dash-dotted line) and $10\,\TeV$ (dashed line).  The effect of the EBL spectrum can be seen in different features in this plot.  The cutoff in $d\Gamma_{\pm}/dE_{\rm EBL}$ above $E_{\rm e, COM}/m_e c^2 \approx 3$ for $E_{\rm ph} = 1\,\TeV$ is due to the cutoff in the EBL spectrum above $10\,\eV$.  The break in $d\Gamma_{\pm}/dE_{\rm EBL}$ at the same position for $E_{\rm ph} = 10\,\TeV$ is due to change in the EBL spectrum at $1\,\eV$. The average COM energies of the produced electrons are $\bar{E}_{\rm e, COM}/m_e c^2 \approx 1.5, 1.7, 2.2,$ and $2.8$ for $E_{\rm ph} = 0.3, 1, 3$, and $10\,\TeV$, respectively.  
 \label{fig:rates}}
\end{figure}

In Figure \ref{fig:rates}, we plot the differential rate of pair production as a function of the COM energy of the electron (and positron), $E_{\rm e,COM}$ for a photon energy of $E_{\rm ph} = 0.3$ (dotted line), $1$ (solid line), $3$ (dash-dotted line) and $10\,\TeV$ (dashed line).  
Note the distribution of COM energy for the electrons (and positrons) depends on the initial photon energy. This is because different energy photons probe different regimes of the EBL spectrum.  Due to the rapid cutoff in the EBL above $10\,\eV$, lower energy VHEGRs produce colder beams.  This is seen in the average COM energies of the produced electrons, which are respectively, $\bar{E}_{\rm e, COM}/m_e c^2 \approx 1.5, 1.7, 2.2$ and $2.8$ for $E_{\rm ph} = 0.3, 1, 3$ and $10\,\TeV$. Hence we expect that these pairs are in the sub-relativistic to mildly-relativistic regimes in their COM frame.

\subsection{Regime of Instability}

Given that the range of $k_BT_b/m_ec^2 = E_{\rm e, COM}/m_e c^2 - 1$ falls between $0.5-2$ for $E_{\rm ph} = 1-10\,\TeV$, we now determine whether or not the reactive or kinetic instabilities apply to these beams. First, it is necessary to determine $n_b/n_t$.  Here the target is the background IGM, so $n_t = \nIGM$, where $\nIGM \approx 2\times 10^{-7}\,(1+\delta)\,(1+z)^3\,\cm^{-3}$ is the mean density of the IGM, $z$ is the redshift, and $\delta$ is the overdensity.  The number density of the TeV beam is more complicated as the production rate of pairs must be balanced against their loss due to plasma instabilities or ICC.  This is discussed extensively in BCP12 and will not be repeated here.  However, we note that the important issue here is the loss rate due to plasma instabilities, which is a nonlinear process.  In BCP12, we assumed that the nonlinear loss rate was the same as the linear growth rate.  This remains to be shown and is the focus of ongoing work, of which this paper lays the initial foundation. 

Still some progress can be made if we use the IC rate as a lower limit to the beam cooling rate. This allows as to get an upper limit on the beam density.The ratio of the beam plasma density to the IGM, $n_b/\nIGM$, is then (BCP12):
\begin{eqnarray}
\frac{\nb}{\nIGM} & \approx 
&
\frac{L_E}{2\pi\Dpp^3\GIC}\frac{1}{\nIGM} \\
&\approx
&
2.3\times10^{-16}
\left(\frac{1+z}{2}\right)^{3\zeta-7}
\left(\frac{E L_E}{10^{45}\erg\,\s^{-1}}\right)
\left(\frac{E}{\TeV}\right)
\,\cm^{-3}\,\nonumber
\label{eq:nbIC}
\end{eqnarray}
at the mean density of the IGM, 
where $L_E$ is the isotropic luminosity per unit energy of the VHEGR source, $E$ is the energy of the VHEGR photon, $\Gamma_{\rm IC}$ is the inverse Compton cooling rate, and
 the mean free path of a VHEGR 
\begin{equation}
\Dpp(E,z) =
35 
\left(\frac{1+z}{2}\right)^{-\zeta}
\left(\frac{E}{1~{\rm TeV}}\right)^{-1}
~{\rm Mpc}\,,
\label{eq:Dpp}
\end{equation}
where  $\zeta=4.5$ for $z<1$ and $\zeta=0$ for $z\ge1$ \citep{Knei_etal:04, Nero-Semi:09}.

In Section \ref{sec:transition}, we derived the controlling parameter that delineates the reactive (Equation \ref{eq:generic reactive regime}) and kinetic regimes (Equation \ref{eq:generic kinetic regime}) by comparing the frequency spread of resonant waves, $\Delta\omega\approx k\Delta v$, with the growth rate, $\Gamma$.  Applying these conditions (eqns.~\ref{eq:generic reactive regime} and \ref{eq:generic kinetic regime}) to the ultrarelativistic $\epm$ pair beams of interest, we find for the controlling parameter:
\begin{multline}
  \frac{\gammabeam^{2/3}}{\sqrt{k_BT_b/m_e c^2}}\left(\frac{n_b}{\nIGM}\right)^{1/3}= 3.2\times 10^{-2} \frac{\gamma_6^{2/3}}{\sqrt{k_BT_b/m_e c^2}}\\
  \qquad\times \left(\frac{1+z}{2}\right)^{\zeta-7/3}
  \left(\frac{E L_E}{10^{45}\erg\,\s^{-1}}\right)^{1/3}
  \left(\frac{E}{\TeV}\right)^{1/3},
  \label{eq:beam oblique condition}
\end{multline}
and
\begin{multline}
  \frac{\gammabeam}{\sqrt{k_BT_b/m_e c^2}}\left(\frac{n_b}{\nIGM}\right)^{1/3} = 3.2 \frac{\gamma_6}{\sqrt{k_BT_b/m_e c^2}}\\
  \qquad\times\left(\frac{1+z}{2}\right)^{\zeta-7/3}
  \left(\frac{E L_E}{10^{45}\erg\,\s^{-1}}\right)^{1/3}
  \left(\frac{E}{\TeV}\right)^{1/3},
  \label{eq:beam bp condition}
\end{multline}
where $\gamma_6 = \gamma/10^6$. 
We see from our reactive (\ref{eq:generic reactive regime}) and kinetic (\ref{eq:generic kinetic regime}) conditions, that the oblique instability always exists in the kinetic regime, but the beam plasma instability is in the reactive regime for $z\gtrsim 1$, for sufficiently cold beams $k_BT_b/m_e c^2 \lesssim 0.5$, which occurs for $E_{\rm ph} \lesssim 0.3\,\TeV$, or for large $\gamma$, which occurs for  $E_{\rm ph} \approx 10\,\TeV$ at $z=0$.

In BCP12, we compared the cold plasma growth rates of the oblique and beam-plasma instabilities and noted that the oblique cold growth rate is larger.  While we also noted that the oblique instability was in the kinetic regime in BCP12, which we confirmed above, we made no effort to study the regime of instability of the beam-plasma case.  Here we have shown that the oblique growth rate is kinetic and the beam-plasma rate is marginally reactive.  This implies that the growth rate of the beam plasma instability is similar to that of the oblique instability.   
In any case, we do not expect that the beam-plasma mode will have a major effect on our earlier results.  First, plasma instabilities losses on the TeV pairs could easily push the beam plasma mode into the kinetic regime by reducing $n_b$, but this requires a proper estimate of the effect of the nonlinear instability.  This is a part of ongoing work and will be presented in a future publication. Second, while it seems that the beam plasma mode may be in the reactive regime, it is not too far from the kinetic regime, i.e., the controlling parameter, $({\gammabeam}/{\sqrt{k_BT_b/m_e c^2}})\left({n_b}/{\nIGM}\right)^{1/3}$, is order unity.  Thus, both the reactive and the kinetic growth rates are similar and it likely makes little difference for the beam plasma mode which regime is assumed (in terms of growth rate).  Therefore, the use of the kinetic growth rate for the oblique mode (and beam-plasma mode) in BCP12 is valid, and the results of this paper buttresses the results of \citet{paperI,paperII,paperIII,paperIV}, and \citet{2015ApJ...811...19L}.

\section{Summary and Conclusion}\label{sec:conclusions}

The ultrarelativistic $\epm$ beams that result from VHEGR-EBL annihilation are subject to powerful plasma beam instabilities including the beam plasma and oblique instability.  In this work, we examined these linear instabilities as they would apply to the ultrarelativistic pair beams.  Our main findings are:
\begin{itemize}
\item{We analytically calculated growth rate of the beam-plasma and oblique instabilities in both the reactive and kinetic regimes.  We have recovered the reactive scalings for the beam-plasma mode  $\Gamma \approx \gamma^{-1}(n_b/n_t)^{1/3}$ and the oblique mode $\Gamma \approx (n_b/\gamma n_t)^{1/3}$.  In the kinetic regime, we have shown that the growth rate for both modes have the same scaling and similar normalization.  Finally, we have shown that the growth rate of the kinetic oblique instability has broad support. Namely, there exists unstable modes that grow at $\approx 0.4\Gamma_0$ for any value of beam temperature and wavevector orientation for relativistic beams.}
\item{We also delineated the regime of applicability of the kinetic and reactive calculation and found, while the kinetic growth rates are similar for both the beam plasma and oblique mode, the condition for transition between the kinetic and reactive regimes are different.  In particular, the beam-plasma mode transitions at a lower value of $\gamma$ in comparison to the oblique mode.  This is due to a difference of $\gamma^{1/3}$ scaling between the two modes. }
\item{We calculate the average COM energy of the ultrarelativistic pair beam using a simplified model of the spectrum of the EBL.  We found that the average energy of these beams range from $E_{\rm e, COM}/m_e c^2 = 1.5-2.8$ for $E_{\rm ph} = 0.3-10\,\TeV$, with colder beams at lower energies.  The average COM energies of the generated pairs implies that the oblique instability is in the kinetic regime, validating our results from BCP12.}
\end{itemize}

\acknowledgements
We thank A. Bret for sharing his notes of the oblique instability and for extensive and enlightening discussions. 
A.E.B.~and M.S.~receive financial support from the Perimeter
Institute for Theoretical Physics and the Natural Sciences and
Engineering Research Council of Canada through a Discovery Grant.
Research at Perimeter Institute is supported by the Government of
Canada through Industry Canada and by the Province of Ontario through
the Ministry of Research and Innovation.
PC is supported by the UWM Research Growth Initiative, the NASA ATP
program through NASA grant NNX13AH43G, and NSF grant AST-1255469.
C.P.~gratefully acknowledges support by the European Research Council under ERC-CoG grant CRAGSMAN-646955 and by the Klaus Tschira Foundation.
E.P.~gratefully acknowledges support by the Kavli Foundation. Support for AL was provided by an Alfred P. Sloan Research Fellowship, NASA ATP Grant NNX14AH35G, and NSF Collaborative Research Grant \#1411920 and CAREER grant \#1455342.  GV acknowledges support from the Australian Research Council, project number DE140101960.

\begin{appendix}

\section{Lorentz Factor Dependence of the Distribution Function and Velocity Dispersion}
\label{sec:lorentz}

Here we explicitly derive the scaling of the parallel and
perpendicular velocity dispersions with the Lorentz factor upon boosting the
distribution function to the lab frame. Let us begin with a distribution
function that is isotropic in the COM frame and depends only on energy.
Therefore in the COM frame, which we denote with the subscript ``COM'', the distribution
function is  $\fCM(E_{\rm COM}(\ppos_{\rm COM},\pmom_{\rm COM}))$.  When we move to the lab
(denoted with subscript ``L'') frame, the integral of the distribution function remains invariant,
i.e., total number, or
\begin{equation}
 N \equiv \int \fLAB \dVLAB = \int \fCM \dVCM.
\end{equation}
It is well-known that under Lorentz transformations \citep{1975ctf..book.....L}, 
\begin{equation}
 \dVLAB = \dVCM,
\end{equation}
so therefore,
\begin{equation}
  \fLAB[\ppos_{\rm L}(\ppos_{\rm COM},\pmom_{\rm COM}),\pmom_{\rm L}(\ppos_{\rm COM},\pmom_{\rm COM})] = \fCM(E_{\rm COM}).
\end{equation}

Now let us consider moments of the distribution function.  For clarity, it is helpful to consider moments of the distribution function first in the COM frame.  The velocity moment is:
\begin{equation}
 \betabarCM = N^{-1}\int \frac {\pmom_{\rm COM}}{\gamma_{\rm COM} m_ec} \fCM \dVCM = 0.
\end{equation}
We consider the lab frame to be  boosted along the x-axis by $\betaBOOST$. More precisely, the initial inertial frame is the COM frame and the lab frame is moving with velocity $\betaBOOST = -|\betaBOOST |$ with respect to the COM frame. This gives:
\begin{equation}\label{eq:betabarLAB}
 \betabarLAB = N^{-1}\int \frac {\pmom_{\rm L}}{\gamma_{\rm L} m_ec} \fLAB \dVLAB  
= N^{-1}\int \frac {\pmom_{\rm L}(\pmom_{\rm COM}, \ppos_{\rm COM})}{\gamma_{\rm L}(\pmom_{\rm COM}, \ppos_{\rm COM}) m_ec} \fCM \dVCM.
\end{equation}
Breaking the components of \betabarLAB\ into components parallel and
perpendicular to the boost, we find: 
\begin{eqnarray}
  \betabarLABpara &=& N^{-1}\int \frac{\betaCMpara +\betaBOOST}{1+\betaBOOST\betaCMpara} \fCM \dVCM \\
  \betabarLABperp &=& N^{-1}\int \frac{\betaCMperp}{\gammabeam\left(1+\betaBOOST\betaCMpara\right)} \fCM \dVCM,
\end{eqnarray}
where $\gammabeam = \left(1-\betaBOOST^2\right)^{-1/2}$ is the Lorentz factor of the boost between the lab and COM frame.  
For $\left|\betaCM\right|,\betaBOOST \ll 1$, we recover the Galilean invariant result, $\betabarLAB \approx \betabarCM + \betaBOOST\hat{\boldsymbol x}$.  
However, this Galilean result no longer holds for relativistic motion.

Now we consider the dispersion around $\betabar$.  In components, the COM frame is:
\begin{equation}
 \dbetabarCMi = N^{-1}\int \beta_i^2 \fCM \dVCM.
\end{equation}
In the lab frame, it is again useful to break it into components -- the parallel component becomes 
\begin{equation}
  \begin{aligned}
    \dbetabarLABpara &= N^{-1}\int (\beta_{\rm L, \parallel}-\betabarLABpara)^2 \fLAB \dVLAB
    = N^{-1}\int \beta_{\rm L,\parallel}^2 \fLAB \dVLAB - \betabarLABpara^2\\
    &= N^{-1}\int \left(\frac{\betaCMpara + \betaBOOST}{1+\betaCMpara\betaBOOST}\right)^2 \fCM \dVCM - \betabarLABpara^2,
  \end{aligned}
  \label{eq:dbetabarLABpara}
\end{equation}
while the perpendicular component becomes
\begin{equation}
  \dbetabarLABperp = N^{-1}\int \beta_{\rm L,\perp}^2 \fLAB \dVLAB
  = N^{-1}\gammabeam^{-2} \int \frac{\betaCMperp^2}{\left(1+\betaCMpara\betaBOOST\right)^2} \fCM \dVCM.
\label{eq:dbetabarLABperp}
\end{equation}

It is easier to look at the perpendicular component first.  It is also more intuitive to study how velocity dispersions scale between the center of mass frame and the lab from for non-relativistic center of mass velocity dispersion.  Hence, for $\left|\betaCM\right| \ll 1$, Equation (\ref{eq:dbetabarLABperp}) becomes to lowest order in $\betaCM$ 
\begin{equation}
 \dbetabarLABperp \approx \frac{\dbetabarCMperp}{\gammabeam^2} \approx \frac {2 k_B T_b}{\gammabeam^2 m_e c^2 }.
\end{equation}
This simple scaling of the perpendicular velocity dispersion can be understood as a scaling with time between two frames boosted relative to each other, where the coordinates perpendicular to the boost axis remain invariant.  This result is also in line with the transformation of temperature as $T\rightarrow T/\gamma$ under a boost, i.e., $mv^2 \approx kT$ -- two factors of $1/\gamma$ from the perpendicular velocity dispersion is countered by one factor of $\gamma$ from the mass.  Let us now consider the parallel component (Equation \ref{eq:dbetabarLABpara}) again to lowest order in $\betaCM$: 
\begin{equation}
 \dbetabarLABpara \approx \frac{\dbetabarCMpara}{\gammabeam^{4}}  \approx \frac {k_B T_b}{\gammabeam^4 m_e c^2 },
\end{equation}
Here, the scaling of the parallel velocity dispersion can be understood as a double scaling of both time and coordinate (along the boost axis) between same two frames boosted relative to each other, giving an extra scaling of $\gamma^{-2}$.
This scaling of the parallel component of the velocity dispersion has important consequences that we explore in the main part of the paper.

\section{Solution for the Reactive Regime}\label{sec:solution reactive}

We begin with the dispersion relation (Equation (\ref{eq:dispersion relation by parts})), which is
\begin{equation}
  \epsilon = 
  1 - \sum_s \frac{m_e\omega_{p,s}^2}{k^2} \int F_s\kvec \cdot \gradp\frac{1}{\omega - \kvec\cdot\vel} d^3p =
  1 - \sum_s \frac{\omega_{p,s}^2}{k^2c^2}\int F_s \frac{k^2c^2 - (\kvec\cdot\vel)^2}{\gamma(\omega - \kvec\cdot\vel)^2} d^3p 
  =
  0\,.
  \label{eq:dispersion relation appendix}
\end{equation}
We then take the limit of Equation (\ref{eq:distribution function beam}) as $k_BT_t  \rightarrow 0$, which yields a $\delta$ function. For the target plasma, we set $v_0=0$, and for a beam plasma $v_0=\vbeam$\footnote{That is, we set $F_s(\bmath{p})=\delta^3(\bmath{p}-\bmath{p}_{0s})$ where $\bmath{p}_{0s}\equiv \gamma_0 m_e v_0 \hat{\bmath{z}}$ is the momentum associated with $v_0$.}.  This leads to Equation (\ref{eq:dispersion relation reactive}), which we reproduce below.  
\begin{equation}\label{eq:dispersion appendix}
 1 - \frac{\omega_{p,t}^2}{\omega^2} - \frac{\omega_{p,b}^2}{\gamma^3(\omega-k_z \vbeam)^2}\frac{\gamma^2 k_x^2 + k_z^2}{k_x^2 + k_z^2} = 0.
\end{equation}
Equation (\ref{eq:dispersion appendix}) can be rewritten as 
\begin{eqnarray}\label{eq:dispersion reactive}
 \left(\omega^2 - \omega_{p,t}^2\right)\left[\left(\omega - k_z \vbeam\right)^2 - \frac{\omega_{p,b}^2}{\gamma^3}\frac{\gamma^2 k_x^2 + k_z^2}{k_x^2 + k_z^2}\right] = \frac{\omega_{p,t}^2\omega_{p,b}^2}{\gamma^3}\frac{\gamma^2 k_x^2 + k_z^2}{k_x^2 + k_z^2},
\end{eqnarray}
where we have added a factor of $\gamma^{-3}\omega_{p,t}^2\omega_{p,b}^2({\gamma^2 k_x^2 + k_z^2})/({k_x^2 + k_z^2})$ to both sides. 
To solve the dispersion relation (\ref{eq:dispersion reactive}), we take $\omega = \omega_{p,t} + \Delta\omega$ and expand to lowest order in
$\Delta\omega$ and $\omega_{p,b}$.  This gives
\begin{equation}
 2\Delta\omega\omega_{p,t}\left(\Delta\omega + \omega_{p,t} - k_z \vbeam\right)^2 = \frac{\omega_{p,t}^2\omega_{p,b}^2}{\gamma^3}\frac{\gamma^2 k_x^2 + k_z^2}{k_x^2 + k_z^2}.
\end{equation}\label{eq:expanded dispersion reactive}
For $\Delta\omega \ll \omega_{p,t} - k_z \vbeam$, $\Delta\omega$ is real and there is no instability.  However, if $k_z = \omega_{p,t}/\vbeam$, we then have
\begin{equation}\label{eq:approximate dispersion reactive}
 \Delta\omega^3 = \omega_{p,t}^3\frac{\omega_{p,b}^2}{2\gamma^3\omega_{p,t}^2}\frac{\gamma^2 Z_x^2 + 1}{Z_x^2 + 1},
\end{equation}
where we have multiplied the fraction on the right hand side by $(\vbeam/\omega_{p,t})^2/(\vbeam/\omega_{p,t})^2$ and $Z_x = k_x\vbeam/\omega_{p,t}$ is the dimensionless wavevector perpendicular to the beam direction.  
Equation (\ref{eq:approximate dispersion reactive}) gives three solutions for $\Delta\omega$: one real and two imaginary (one growing and one damping).  The maximum growth rate is then
\begin{equation}\label{eq:growth rate reactive appendix}
 \Gamma = \frac{\sqrt{3}}{2^{4/3}}\left(\frac{n_b}{n_t}\right)^{1/3}\left(\frac{\gamma^2 Z_x^2 + 1}{Z_x^2 + 1}\right)^{1/3}\frac{\omega_{p,t}}{\gamma}
\end{equation}

\section{Solution for the Kinetic Regime}\label{sec:solution kinetic}

To find the growth rate for the kinetic regime, we begin first with the distribution function for the target plasma
\begin{equation}
 F_t = \left(\frac{1}{2\pi m_e k_B T_t}\right)^{3/2}\exp\left(-\frac{p^2}{2m_e k_B T_t}\right).
\end{equation}
We assume that the target plasma is nonrelativistic with a momentum $\bmath{p}=m_e\vel$,  and $T_t$ is the temperature of the target background plasma.
For the beam plasma, we adopt the Maxwell-J{\"u}ttner distribution (Equation~\ref{eq:distribution function beam}).
Inserting these into the dispersion relation (Equation \ref{eq:dispersion relation}), we find
\begin{equation}\label{eq:ap dispersion kinetic}
 1 - \frac{\omega_{p,t}^2}{k^2c^2}\int F_t\frac{k^2c^2 - (\kvec\cdot\vel)^2}{\gamma(\omega - \kvec\cdot\vel)^2} d^3p + \frac{m_e\omega_{p,b}^2}{k^2}\int \frac{\kvec\cdot\gradp F_b}{\omega - \kvec\cdot\vel}d^3p  = 0,
\end{equation}
where we have integrated by parts only the second term, associated with the target plasma. 

As the target plasma is nonrelativistic, we can take $v\ll c$ and $\gamma \rightarrow 1$.  Expanding the denominator in powers of $v$, we find\footnote{An alert reader will note that Lorentz factor, $\gamma$, and the second term in the numerator both contribute to the expansion in powers of $v$ at second order.  These contributions are the result of the minor deviations from the Lorentz factor of the nonrelativistic electrons and the subtle different between momentum and velocity at order $v^2/c^2$.  These corrections correct the plasma frequency, $\omega_p$ at order $v^2/c^2$, but do not change the physics of the oscillations, i.e., they are independent of the wavevector.  Hence, we ignore these effects while keeping the $O(v^2/c^2)$ correction that determine the Langmuir wave because these corrections depend on the wavevector.}
\begin{equation}
  \int F_t\frac{k^2c^2 - (\kvec\cdot\vel)^2}{\gamma(\omega - \kvec\cdot\vel)^2} d^3p
  \approx k^2c^2  \int F_t \left(\frac {1}{\omega^2} + \frac {2\kvec\cdot\vel}{\omega^3} + \frac{3 (\kvec\cdot\vel)^2}{\omega^4}\right) d^3p
  \approx \frac {k^2c^2}{\omega^2}\left(1 + 3 {k^2\lambda_D^2}\right),
\label{eq:dispersion-real-part}
\end{equation}
where the second term is zero because it is odd, $\lambda_D^2 = k_B T_t/m_e \omega_p^2$ is the Debye length, and we have assumed that $k^2\lambda_D^2 \ll 1$ and $\omega \approx \omega_p$ in the last term on the RHS\footnote{A direct solution to Equation (\ref{eq:dispersion-real-part}) without approximating $\omega\approx \omega_p$ will reveal waves with nontrivial growth or damping rates.  These wave are not legitimate and result from the Taylor expansion of the denominator of Equation (\ref{eq:dispersion-real-part}).  A correct treatment of Equation (\ref{eq:dispersion-real-part}) with the appropriate Landau contours will give the correct growing or damping behavoirs for waves with phase speeds approximately that of the electron phase speeds.}. If we ignore the third term in the kinetic dispersion relation (\ref{eq:dispersion kinetic}), this yields two plasma modes: an undamped plasma oscillation mode with $\omega = \omega_{p,t}$ and a longitudinal electron plasma wave, i.e., Langmuir wave, with
\begin{equation}
\omega \approx \omega_{p,t}\left(1 + \frac 3 2 k^2\lambda_{D,t}^2\right).
\end{equation}

To compute the contribution from the beam term, we will reorient our coordinate system and define the $z'$-axis along the wavevector, $\kvec$.  In this case we have the beam taking on a non-$z'$ component,  $\vel_b = v_{bz'}\zphat+v_{bx'}\xphat$.   
This frame moves with a velocity, $\vel_{\rm ph} = \omega_k/k \zphat$.  With an eye toward computing the residue that will appear in Equation (\ref{eq:dispersion kinetic}), we define $p_{z'}=\gz\vz E_\perp$, $E=\gz E_\perp$, and
$E_\perp=\sqrt{m^2+p_\perp^2}$ is the perpendicular energy.  In this case, we can rewrite the beam distribution function as
\begin{eqnarray}
F_B &=& \frac{m_e c^2}{4\pi\gammabeam k_B T_b K_2(m_ec^2/k_B T_b )m_e^3c^3} \exp\left(-\frac {\gammabeam(E - v_{b,z'} p_{z'} - v_{b,x'} p_{x'})} {k_B T_b}\right)\nonumber\\
&=& \frac{\mu m_e^{-3}c^{-3}}{4\pi\gammabeam K_2(\mu)}  \exp\left(-\frac {\gammabeam\gz(c^2 - v_{b,z'}v_{z'})E_\perp)} {k_BT_b c^2}\right)\exp\left(\frac{\gammabeam v_{b,x'} p_{x'}}{k_BT_b}\right),\label{eq:dist2}
\end{eqnarray}
where we define $\mu = m_ec^2/k_BT_b$.
Inserting the equation into (\ref{eq:ap dispersion kinetic}) and using the results of Equation (\ref{eq:dispersion-real-part}), we find 
\begin{equation}
 1 - \frac{\omega_{p,t}^2}{\omega^2}\left(1 + 3 {k^2\lambda_D^2}\right) + i\frac{\pi n_b}{n_t}   m_e v_B^2 R = 0, 
\end{equation}
which involves the integral of 
\begin{equation}
R\equiv \int \frac{k\partial F_b/\partial p_{z'}}{\omega - k_{z'}v_{bz'}}d^3p,
\end{equation}
where $R$ is the residue for $p_{z'}$ such that $v_{z'} = v_{\rm ph}$.  

We can assume that $k\lambda_D \ll 1$ as the thermal velocity of the background plasma is much smaller than the speed of the ultrarelativistic beam.  We then take $\omega = \omega_r + i\Gamma$, where $\omega_r = \Re(\omega)$ is the real part of $\omega$ and the growth rate $\Gamma \ll \omega_r$, to find:
\begin{equation}
\Gamma \approx
-\omega_p\frac{\pi n_b}{2 n_t}   m_e v_B^2 R.
\end{equation}

Here two elements contribute to the pole:
\begin{equation}
\left.
\frac{\partial}{\partial p_{z'}} \left(\omega-k v_{bz'}\right)
\right|_{\rm pole}
=
\left.
- k \left(
  \frac{c^2}{E} - \frac{p_{z'}^2 c^4}{E^3}
\right)
\right|_{\rm pole}
=
- \frac{k c^2}{\gph^3 E_\perp}\,,
\end{equation}
and 
\begin{equation}
\begin{aligned}
\left.
k\frac{\partial F_b}{\partial p_{z'}}
\right|_{\rm pole}
&=
\left.
-\frac{k\gammabeam}{k_BT_b} \left( \frac{p_{z'} c^2}{E} - v_{b,z'}\right)
F_b
\right|_{\rm pole} \\
&=
-
\frac{
  k
  (\vph-v_{b,z'}) \mu^2}{
  4\pi m_e^{4}c^{5}K_2(\mu)
} \exp\left(-\frac {\gammabeam\gph(c^2 - v_{b,z'}v_{\rm ph})E_\perp)} {k_BT_b c^2}\right)\exp\left(\frac{\gammabeam v_{b,x'} p_{x'}}{k_BT_b}\right).
\end{aligned}
\end{equation}

Putting this all together, the residue is
\begin{equation}
R
=
\frac{
  \gph^3 (\vph-v_{b,z'})\mu^2}{4\pi K_2(\mu)
}
\frac{\kI}{m_e^4c^7}
\quad\text{where}\quad
\kI \equiv \int d^2\!p_\perp E_\perp \exp\left(-\frac{\cG(E_\perp-w p_x)}{k_B T_b}\right),
\label{eq:Rogen I-integral}
\end{equation}
$\cG\equiv\gammabeam\gph(1-v_{b,z'}\vph/c^2)$, and
$w\equiv \gammabeam v_{b,x'}/\cG \le 1$.  This latter inequality is guaranteed as 
\begin{equation}
\cG E_\perp - \gammabeam v_{b,x'} p_{x'} = \gammabeam(E - v_{b,z'} p_{z'} - v_{b,x'}p_{x'})|_{v_{b,z'}=v_{\rm ph}} > 0
\end{equation}
is the energy in beam frame and is therefore positive definite.  Noting that the $\exp(-wp_{x'})$ term appears as a boosted distribution, we boost by $w$
along the $x'$-axis, removing the anisotropic term from the
exponential.

Thus, we define $p_{x'}'=\gamma_w(p_{x'}-w E_\perp/c^2)$ and $p_{y'}'=p_{y'}$ and find:
\begin{equation}
  E_\perp = \gamma_w(E'_\perp + w p'_x)
  \quad\text{and}\quad
  d p_x d p_y = \frac{E_\perp}{E'_\perp} d p'_x d p'_y.
\end{equation}
Inserting this into Equation (\ref{eq:Rogen I-integral}), we find
\begin{equation}
\begin{aligned}
\kI &= 
\int d^2\!p'_\perp \frac{E_\perp^2}{E'_\perp} \exp\left(-\frac{\cG'E'_\perp}{k_B T_b}\right) 
=
\pi \gamma_w^2 \int_0^\infty dp'^2_\perp \left(
  E'_\perp + \frac{w^2}{2} \frac{p'^2_\perp}{E'_\perp}
\right) \exp\left(-\frac{\cG'E'_\perp}{k_B T_b}\right)\,,
\end{aligned}
\end{equation}
where $\cG'\equiv\cG/\gamma_w$. Note in the second line that we have used isotropy in $\bp_{\perp}'$ to eliminate terms linear in $\bp_{\perp}'$. 
Using the following integrals: 
\begin{equation}
  \int_0^\infty dx \sqrt{1+x} e^{-a\sqrt{1+x}} = \frac{2 e^{-a}}{a^3}\left(a^2 + 2a + 2\right)
  \quad\text{and}\quad
\int_0^\infty dx \frac{x}{\sqrt{1+x}} e^{-a\sqrt{1+x}}
=
\frac{4(a+1)}{a^3} e^{-a}\,,
\end{equation}
we find
\begin{equation}
\kI =
\frac{2 \pi \gamma_w^2 m_e^3c^4}{\cG'^3\mu^3} \left[
\left(\cG'^2\mu^2+2\cG'\mu+2\right)
+
\frac{w^2}{2c^2}
\left(2\cG'\mu+2\right)
\right]\exp\left(-\cG'\mu\right).
\end{equation}
Inserting this into (\ref{eq:Rogen I-integral}) yields
\begin{equation}
R =
\frac{\gph^3 \gamma_w^2 (\vph-v_{b,z'})}{ 2 \mu  \cG'^3 K_2(\mu) c} 
\left[
\left(\cG'^2\mu^2 + 2\cG'\mu + 2\right)
+
\frac{\gammabeam^2 v_{b,x'}^2}{2\cG'^2 c^2} \left(2 \cG'\mu + 2\right)
\right] \exp\left(-\cG'\mu\right)
\end{equation}
and therefore,
\begin{equation}
\Gamma \approx
- \Gamma_0
\frac{\pi \gamma_w^2 \gph^3 (\vph-v_{b,z'})}{4 \gammabeam \mu^2 K_2(\mu) \cG'^3 c}
\left[
\left( \cG'^2\mu^2 + 2\cG'\mu + 2 \right) 
+
\frac{\gammabeam^2 v_{b,x'}^2}{2\cG'^2c^2} \left(2 \cG'\mu+ 2\right)
\right]
\exp(-\cG'\mu),
\label{eq:OGgen}
\end{equation}
where $\Gamma_0 \equiv \omega_p \gammabeam (n_b/n_t) (m_e v_B^2/k_B T_b)$ is the typical maximum growth rate.

\end{appendix}

\bibliography{ms}
\bibliographystyle{apj}

\end{document}